# Ergodic Theory and the Structure of Noncommutative Space-Time

by

James Moffat, Teodora Oniga, Charles H.-T. Wang

*Department of Physics, University of Aberdeen, King's College, Aberdeen AB24 3UE, UK*

**Abstract**

We develop further our fibre bundle construct of non-commutative space-time on a Minkowski base space. We assume space-time is non-commutative due to the existence of additional non-commutative algebraic structure at each point **x** of space-time, forming a quantum operator 'fibre algebra' **A(x)**. This structure then corresponds to the single fibre of a fibre bundle. A gauge group acts on each fibre algebra locally, while a 'section' through this bundle is then a quantum field of the form $\{A(x); x \in M\}$ with $M$ the underlying space-time manifold. In addition, we assume a local algebra $O(D)$ corresponding to the algebra of sections of such a principal fibre bundle with base space a finite and bounded subset of space-time, $D$. The algebraic operations of addition and multiplication are assumed defined fibrewise for this algebra of sections.

We characterise 'ergodic' extremal quantum states of the fibre algebra invariant under the subgroup $T$ of local translations of space-time of the Poincare group $P$ in terms of a non-commutative extension of entropy applied to the subgroup $T$. We also characterise the existence of $T$- invariant states by generalizing to the non-commutative case Kakutani's work on wandering projections. This leads on to a classification of the structure of the local algebra $O(D)$ by using a '$T$-Twisted' equivalence relation, including a full analysis of the $T$-type III case. In particular we show that $O(D)$ is $T$-type III if and only if the crossed product algebra $O(D) \times T$ is type III in the sense of Murray-von Neumann.

## Introduction

*Ergodic Theory in the Classical (Commutative) Case*

In the commutative case, a general von Neumann algebra $R$ is isomorphic to the set $L^\infty(Z, \nu)$ of essentially bounded, measurable, complex-valued functions on the locally compact set $Z$ with $\nu$ a positive regular borel measure. If $G$ is a group of automorphism of $R$ then $G$ is by definition isomorphic to a group of automorphisms of $L^\infty(Z,\nu)$ which we also denote as $G$. Projections $P$ in the algebra $R$ become, under the isomorphism, characteristic functions of borel subsets of $Z$. If $g: P \to Q$ and $P$ is isomorphic to $\chi_E$ for the borel set $E$ then $Q$ is also a projection thus is isomorphic to $\chi_F$ for some borel subset $F$. Thus the group $G$ induces a group of transformations of the $\sigma$-ring $\mathcal{B}$ of borel subsets of $Z$.

Let $T$ be such a transformation which is measure preserving, then for any borel set $E$ in $\mathcal{B}$, $\nu(T^{-1}(E)) = \nu(E)$ i.e. the measure $\nu$ is $T$-invariant. $T$ is defined to be ergodic if it mixes the space. i.e. $T^{-1}(E) = E$ modulo a null set implies either $\nu(E) = 0$ or $\nu(Z \setminus E) = 0$. If $T$ is ergodic in this sense and the measure $\nu$ is $T$-invariant, then $\nu$ is defined to be an ergodic measure (Halmos, 1956). It then follows that if $Z$ is compact, $\nu$ is a probability measure on $Z$ which is ergodic, and $T$ and its inverse are continuous mappings, then this is equivalent to $\nu$ being an extreme point of the invariant measures on $Z$. For if $\mathcal{E}$ denotes the set of such invariant probability measures, and $\nu$ is an extreme point of $\mathcal{E}$, then any measurable set $E$ with $0 < \lambda = \nu(E) < 1$ allows the construction of measures

$\mu_1(*) = \frac{1}{\lambda} \nu(* \cap E)$ and $\mu_2(*) = \frac{1}{1-\lambda} \nu(* \cap (Z \setminus E))$ such that $\nu = \lambda \mu_1 + (1-\lambda)\mu_2$; a contradiction.

Conversely if $\nu$ is a probability measure on $Z$ which is ergodic, and $0 < \mu < \nu$, then we have, for any Borel set $A$, $\mu(A) = \int_A f(x) d\nu(x)$ for some $f \in L^1(Z,\nu)$ and $f$ is a $T$-invariant, positive function by the properties of probability measures. If $f$ is not constant we can define Borel sets $S_1 = \{x \in Z; f(x) < t\}$ and $S_2 = \{x \in Z; f(x) > t\}$ for some positive real $t$. Both sets are invariant and non-trivial, thus they must both have measure 1; a contradiction. Hence $f$ is constant and the measure is an extreme point since for any convex combination of measures, $\nu$ dominates these measures and this leads to the tautology $\nu = \lambda\nu + (1-\lambda)\nu$ for some $\lambda: 0 < \lambda < 1$.

Given a partition $p$ of $Z$ into measurable subsets $\{A_j; j = 1, 2, 3, \ldots\}$, If $Z$ is compact and has total measure equal to 1 then we can interpret the value $\mu(A_j)$ as the probability of the set $A_j$. The expression $-\log \mu(A_j)$ is then a measure of the description length or Kolmogorov Complexity of the partition subset $A_j$. A measure preserving transformation such as $T$ transforms the partition $p$ into the partition $Tp$ which is $\{TA_j; j=1,2,3....\}$. The entropy or, equivalently, the expected value of the Kolmogorov Complexity of the partition $p$ is defined as $-\sum_j \mu(A_j)\log\mu(A_j)$. Since $T$ is measure preserving, the partitions $p$ and $Tp$ have the same entropy.

(Hopf, 1932) was interested in the question of when a measure representable as a Lebesgue integral is invariant under a measurable transformation $T$. Hopf considers a partition of the measurable set into subsets and the effect of multiples $T^k$ of $T$ acting on those subsets. In operator theoretic language we can express this as follows (Stormer 1973). We replace measurable sets by projections and the group $\{T^k; k \in Z\}$ by a general discrete group $G$ of automorphisms of a *commutative* von Neumann algebra $R$ acting on a Hilbert space $H$ which is implemented by the unitary representation $g \to U_g$ from $G$ to the set of unitaries acting on $H$. Two projections $H$ and $K$ in $R$ are *Hopf equivalent* if there is an orthogonal family of projections $E_j$ in $R$ and group elements $g_j \in G$ with $H = \sum_j E_j$ and $K = \sum_j U_{g_j}^* E_j U_{g_j}$. This equivalence leads to a partition based criterion, '$H$-finiteness', for the existence of a finite invariant measure. For this kind of orthogonal partition we can define the entropy as being derived from the partition weightings (all equal in this case). With this definition it is then clear that if the two projections $H$ and $K$ in $R$ are Hopf equivalent then they have the same entropy relative to an invariant measure. This idea is easily extended to the noncommutative case, as we will see later.

*Noncommutative (Quantum) Ergodic Theory*

The low energy regime beyond the standard model can be represented, we postulate, by linearised gravity with matter on a flat space-time manifold $M$. For this regime, invariance of quantum states to the Poincare group is a key symmetry and new results were presented in (Moffat, Oniga and Wang, 2016). Within the resulting fibre bundle construct defined there, our focus is on the local fibre algebra **A(x)** and the subgroup $T$ of the Poincare group consisting of translations of space-time as a group of automorphisms of **A(x)**.

If **A(x)** is a fibre algebra and the group $T$ of translations of space-time a subgroup of the Poincare group, we define $\alpha: g \to \alpha_g$ to be a representation of $T$ as automorphisms of **A(x)**. Let $f$ be a faithful normal state of **A(x)** and define the 'induced' transformation
$v_g(f) = f \circ \alpha_g$. This means that $v_g(f)$ is also a normal state and $v_g(f)(A) = f(\alpha_g(A))$. Since the subgroup $T$ is abelian, and the mapping $g \to v(g)$ is a group homomorphism, the set $\{v(g); g \in T\}$ is a continuous group of commuting transformations of the dual space **A(x)***. If $f$ is a state of the algebra then define $\mathcal{E}$ to be the weak* closed convex hull of the set $\{v(g)f; g \in T\}$. Then $\mathcal{E}$ is a weak* compact convex set and each $v(g): E \to E$. By the Markov-Kakutani fixed point theorem (Reed and Simon, 1980) it follows that $\mathcal{E}$ has an invariant element. In other words, the group $T$ has the fixed point property (Piers, 1984) and thus is amenable. In summary, because $T$ is an abelian group and locally compact it is an amenable group, since the closed convex hull of any quantum state of the system contains a $T$-invariant state. This leads us to define the following;

**Definition.** Let **A(x)** be a fibre algebra and the group $T$ of translations of space-time a subgroup of the Poincare group. Let $\alpha: g \to \alpha_g$ be a representation of $T$ as automorphisms of **A(x)**. The group representation $\alpha$ acts ergodically on **A(x)** if given a projection $E$ in **A(x)**, $\alpha_g(E) = E \: \forall g \in T$ implies that $E=0$ or $E=I$.

This definition is a direct generalisation of the commutative case where **A(x)** is the set of essentially bounded measurable functions on a locally compact space with a regular borel measure, discussed above. We can also have the following non-commutative generalisation of an ergodic probability measure as an extreme point of the set of invariant measures; as first pointed out by (Segal, 1951).

**Definition.** Let **A(x)** be a fibre algebra and $\alpha: g \to \alpha_g$ a representation of the translation subgroup $T$ of the Poincare group $P$ as automorphisms of **A(x)**. A quantum state $f$ of A(x) is $\alpha$-invariant if $f(\alpha_g(A)) = f(A) \quad \forall A$ in **A(x).** If $f$ is a normal (i.e. density matrix) state and an extreme point of the set of invariant states, then $f$ is defined to be a $T$-ergodic state.

Note that the set of invariant states is a compact convex subset of the quantum state space of **A(x)** and is thus generated by its extreme points (Krein and Milman, 1940). There is a non-trivial invariant state for the amenable group $T$, as discussed earlier, thus there is an extremal $T$- invariant state of **A(x)**. Since, by definition, the Hilbert space representation on which **A(x)** acts is separable, the algebra contains a faithful normal state and hence a *T*-invariant normal state. The norm limit of a set of normal states is again normal, (Kadison and Ringrose, 1983) and thus for a separable Hilbert space, the fibre algebra **A(x)** with the assumptions above always contains a *T*-ergodic state.

**Definition.** The fibre algebra **A(x)** is a quantum operator algebra and thus always contains an identity operator *I*. If *f* is a quantum state of **A(x)** then by definition, *f(I)=1*. The support of *f* is the unique smallest projection *E* in **A(x)** such that *f(E) =1* and is denoted $E_f$.

We also require the following definition;

**Definition.** E is an α-invariant projection in **A(x)** if $\alpha_g(E) = E \quad \forall g \in T$.

The following result is well known in the finite dimensional (matrix algebra) case. For completeness we give a general proof.

**Theorem 2**. Let **A(x)** be a fibre algebra and $\alpha: g \to \alpha_g$ a representation of the translation subgroup *T* of the Poincare group *P* as (gauge) automorphisms of **A(x)**. Assume there exists at least one state *f* of **A(x)** which is α-invariant. Then the support of *f*, $E_f$, is an invariant projection and *f* is a normal and ergodic state if and only if the representation $\alpha$ acts ergodically on the cut down algebra $E_f \mathbf{A(x)} E_f$.

**Proof.** We start with the observation that assuming *f* is α-invariant implies that $f(\alpha_g(E_f)) = f(E_f) = 1 \quad \forall g \in T$; we call such states 'symmetric'. By uniqueness of the support of *f* it follows that $E_f$ is an α-invariant projection. Let $\pi$ be the Gelfand-Naimark-Segal (GNS) representation of **A(x)** induced by the state *f* on the Hilbert space *H(f)*. We make the simplifying assumption for now that *f* is a faithful state; i.e. $E_f = I$, and revisit this assumption later. In this case the von Neumann algebra π(**A(x)**) has a separating-generating vector $\xi$ and the representation $\pi$ is a *-isomorphism. Define the unitary group $U_g \pi(A)\xi = \pi(\alpha_g(A))\xi$ on a dense subset of *H(f)*, then $U_g$ extends to a unitary on $H(f) = \{\pi(B)\xi; B \text{ in the fibre algebra}\}^-$ where $\{\}^-$ denotes closure of the set in the norm topology. The mapping $U: g \to U_g$ is then a unitary representation of the translation group T and for B a quantum observable in the fibre algebra **A(x)** we have $U_g \pi(B) U_g^* = \pi(\alpha_g(B))$ $\forall g \in T$ i.e. the unitary representation U implements the automorphic representation $\alpha : g \to \alpha_g$.

Consider now the involution mapping on π(**A(x)**) defined as $A \to A^*$. This induces an anti-linear mapping on a dense subset of the Hilbert space H(f); $S : A\xi \to A^*\xi$. Moreover, this extends to a mapping with closed graph which we also denote by S. By the theorem of Tomita-Takesaki (Kadison and Ringrose, 1983) S has a polar decomposition $S = J\Delta^{\frac{1}{2}}$ such that J(**A(x)**)J = π(**A(x)**)'; the commutant of the fibre algebra π(A(x)). If $x = B\xi$ is in the domain of S, then it follows that $U_g x$ also lies in the domain of S, and we have the relationship;

$$U_g SB\xi = U_g B^*\xi = \alpha_g(B^*)\xi = \alpha_g(B)^*\xi = S\alpha_g(B)\xi = SU_g B\xi$$

This leads to the conclusion that, on the domain of S, we have $U_g S = S U_g$.

Then we have;

$$S = U_g S U_g^* = U_g J \Delta^{\frac{1}{2}} U_g^* = U_g J U_g^* U_g \Delta^{\frac{1}{2}} U_g^*$$

By uniqueness of the polar decomposition, $J = U_g J U_g^*$; J and $U_g$ commute for all $g \in T$. From this we deduce that;

$B \in \pi(A(x)) \cap \{U_g ; g \in T\}'$ implies that $JBJU_g = U_g JBJ$ for all $g \in T$.
Thus $JBJ \in \{U_g ; g \in T\}' \cap \pi(A(x))'$.
Conversely, if $C \in \{U_g ; g \in T\}' \cap \pi(A(x))'$ then $C = JBJ$ for some $B \in \pi(A(x))$ and $JBJU_g = U_g JBJ$ implies $JBU_g J = JU_g BJ$ and thus $BU_g = U_g B$ so that $B \in \{U_g ; g \in T\}'$.

We conclude that $J\{\pi(A(x)) \cap \{U_g ; g \in T\}'\}J = \pi(A(x))' \cap \{U_g ; g \in T\}'$

The automorphic representation $\alpha: g \to \alpha_g$ of $T$ acts ergodically if and only if $\pi(A(x)) \cap \{U_g; g \in T\}'$ is trivial, containing only the projections 0 and I and thus consisting of the set of complex multiples of *I*. From the reasoning above it follows that the representation $\alpha: g \to \alpha_g$ of $T$ acts ergodically if and only if $\pi(A(x))' \cap \{U_g; g \in T\}'$ invariant is also trivial.

If *E* is a projection in the set $\pi(A(x))' \cap \{U_g; g \in T\}'$ we can define a state $f_E(A) = \frac{\langle \xi, E\pi(A)\xi \rangle}{\langle \xi, E\xi \rangle}$ on the fibre algebra **A(x)**. Then $f_E = \omega_{E\xi} \circ \pi$ is a state dominated by $f = \omega_\xi \circ \pi$ and we have;

$$f(A) = \frac{\langle \xi, \pi(A)\xi \rangle}{\langle \xi, \xi \rangle} = \lambda \frac{\langle \xi, E\pi(A)\xi \rangle}{\langle \xi, E\xi \rangle} + (1-\lambda) \frac{\langle \xi, (I-E)\pi(A)\xi \rangle}{\langle \xi, (I-E)\xi \rangle} \text{ for } A \in \mathbf{A(x)},$$

where $\lambda = \frac{\langle \xi, E\xi \rangle}{\langle \xi, \xi \rangle} = \frac{\|E\xi\|^2}{\|\xi\|^2}$ and $1 - \lambda = \frac{\|\xi\|^2 - \|E\xi\|^2}{\|\xi\|^2} = \frac{\langle \xi, (I-E)\xi \rangle}{\|\xi\|^2}$

Thus *f* is an extremal invariant state if and only if the projection *E=0 or I*. The result follows for the support of *f* equal to 1. Finally, we need to extend the result to a general invariant state *f* with support $E_f, 0 < E_f < I$. This follows from what we have already proved, since the restriction of *f* to $E_f \mathbf{A(x)} E_f$ is a faithful state, and a state extremal among the invariant states of the cut down algebra $E_f \mathbf{A(x)} E_f$ is also extremal among the invariant states of the full fibre algebra **A(x).** This follows from the fact that if *f* is a convex combination of states from the full fibre algebra, then each of them has a support less than or equal to $E_f$. In the next section we develop and prove a noncommutative version of a well-known result in classical ergodic theory and use it to characterise the existence of such symmetric states.

**Wandering Projections and Invariant Symmetry States**

(Hajian and Kakutani, 1964) defined a wandering set as follows;

**Definition.** Let $(X, B, \mu)$ be a measure space with finite measure; $\mu(X) < \infty$ and where *B* is the set of all measurable subsets of *X*. Let *T* be a bijective transformation of *X* such that both *T* and its inverse are measurable mappings. A wandering set for *T* is a measureable subset *S* of *X* such that the sets $\{T^{nk}(S)\}$ are disjoint, for some infinite sequence of integers *nk*.

**Definition.** Two measures $\nu$ and $\mu$ on the measure space X are said to be equivalent if they share the same null sets. A measure $\nu$ is *T*-invariant if $\nu(T(E)) = \nu(E)$ for all measurable subsets *E* of *X*.

With these definitions, (Hajian and Kakutani, 1964) showed that there is a finite *T*-invariant measure $\nu$ on *X*, equivalent to $\mu$, if and only if there are no wandering subsets of *X*.

If now we consider an abelian von Neumann algebra *R*, then *R* is isomorphic to *C (X)* with *X* a compact stonean space of finite measure, and the positive, normal, regular borel measures on *X* correspond to the normal states of *R*. By (Dixmier, 1951) we can characterise these normal measures as being equivalent to measures which annihilate each nowhere dense subset of *X*. It follows that if measures $\nu$ and $\mu$ on the measure space *X* are equivalent and measure $\nu$ is normal, then measure $\mu$ is also normal.

If $\theta$ is a continuous automorphism of the abelian algebra *R*, isomorphic to *C (X)*, then we can define the homeomorphism *T* of *X* by $\theta f(x) = f(Tx)$ for $f \in C(X)$. By the result quoted above, if $\mu$ is a normal measure on *X* with support equal to *X*, there is a measure equivalent to $\mu$ which is *T*-invariant if and only if there are no wandering measurable subsets *E* of *X*.

If such a set *E* did exist, such that the sets $\{T^{nk}(E)\}$ are disjoint, for some infinite sequence of integers *nk*, then by regularity of $\mu$ we can assume that *E* is closed. Since *X* is a stonean space, *E* is both open and closed. Thus the characteristic function $\chi_E$ corresponds to a projection in the algebra *R* and the set of projections $\theta^{nk}(\chi_E)$ is an orthogonal set. From the algebraic perspective then we can say the following. Given an abelian von Neumann algebra *R*, an automorphism $\theta$ of *R* and a faithful normal state acting on *R*. Then there is a faithful normal $\theta$-invariant state acting on *R* if and only if there are no non-trivial projections *E* in *R* such that for some infinite sequence of integers *nk*, the projections $\theta^{nk}(E)$ are mutually orthogonal. It can be easily shown that for a commutative algebra this condition on the set of projections is equivalent to the requirement that that there are no nonzero projections *E* with $\theta^{nk}(E) \to 0$ in the ultraweak topology as $nk \to \infty$ for some infinite sequence *nk* of integers. This new formulation now generalises easily to the noncommutative (quantum) case as follows.

**Definition.** Let **R** be a von Neumann algebra, *G* a group of automorphisms of **R**. Then a nontrivial projection *E* in **R** is wandering if *E* is such that; $g_{n_k}(E) \to 0$ for some infinite sequence $g_{n_k}$ in *G*. Convergence is defined in the weak operator topology.

If **A(x)** is a fibre algebra then it is a von Neumann algebra with trivial centre and is countably decomposable. Let $\alpha: g \to \alpha_g$ be a group representation of the translation subgroup *T* of the Poincare group which is ultraweakly continuous.

**Theorem 3.** There is a faithful normal translation invariant quantum state on the fibre algebra **A(x)** if and only if there are no wandering projections in **A(x)**.

Clearly if *E* is a projection in **A(x)** such that $\alpha_{g_{n_k}}(E) \to 0$ for some infinite sequence $g_{n_k}$ in *G* and *f* is a faithful, normal *α*-invariant state, then *f(E)* = 0, thus *E*=0.

The proof of the converse is based on work by M Takesaki on singular states (Takesaki, 1959).

We assume that there are no wandering projections in **A(x).** The fibre algebra **A(x)** has a faithful normal state *f*. By the fixed point property, applied to the set; $\mathscr{E}$ =weak* closed convex hull of $\{v(g)f; g \in T\}$, **A(x)** has an invariant state which we denote as *h*. We need to show that *h* is both normal and faithful. By (Takesaki, 1958) *h* has a unique decomposition $h = h_n + h_s$ with $h_n$ a normal positive linear functional and $h_s$ a singular positive linear functional. By uniqueness of the decomposition, both of these linear functionals are also *α*-invariant. Let *S* be the support of $h_n$ so that $0 \leq S \leq I$. If $S \neq I$ we can choose a projection *F* with $0 < F < I - E$ and $h_s(F) = 0$ (Takesaki, 1959).

Let $\lambda = \inf_{g \in T}(f \circ \alpha_g(F))$. Since $h = h_n + h_s$, we have *h(F)* = *0*. Therefore *λ* = 0. Thus there is a sequence $g_{n_k}$ with $f \circ \alpha_{g_{n_k}}(F) \to 0$. Since *f* is faithful and normal this implies that $\alpha_{g_{n_k}}(F) \to 0$ in the weak operator topology; i.e. *F* is a wandering projection. This contradiction shows that the support of $h_n$ equals I and $h_n$ is the required normal, faithful invariant state.

**The Structure of the Local Algebra *O(D)***

For each event point **x** in Minkowski space-time, we have a fibre algebra **A(x)** defined as a von Neumann algebra with trivial centre and a faithful representation as an algebra of operators acting on a separable Hilbert space. Thus *O(D)* is an associative principal fibre subbundle; *associative* in the sense that a Lie group (the translation subgroup of the Poincare group) acts on each fibre; a *subbundle* in the sense that only that subset of $\{\mathbf{A}(\mathbf{x}); \mathbf{x} \in M\}$ with $\mathbf{x} \in D$ is of physical interest.

The local von Neumann algebra *O(D)* does not necessarily have a trivial centre; its structure is more complex in some ways. We assume that the quantum system it represents has an energy operator with discrete countable eigenstates and we thus assume also that *O(D)* is separable. We propose to use the ideas of noncommutative ergodic theory to gain insight into the structure of *O(D)*, as we now describe.

Von Neumann introduced the idea of equivalence of measurement 'projection' operators as a way of gaining traction on the structure of a general von Neumann algebra, see for example (Kadison and Ringrose, 1983). Much of this analysis centres around the question of whether or not the algebra possesses a finite trace, extending the idea of the trace of a finite matrix operator as the sum of its observable eigenvalues. This analysis was enhanced to take account of groups of (unitarily implemented) automorphisms of the algebra by (Stormer, 1973). This allows him to define a 'G-equivalence' of projections which generalises to the non-commutative quantum case the definition (Hopf, 1932) used in standard commutative ergodic theory. One of us (Moffat, 1974), extended this work to a characterisation of the tensor product of 'G-type III' algebras. As a result of this previous work we can now develop a classification of the structure of our local algebra *O(D)*. We do this by applying these earlier results where the group concerned is now the subgroup *T* of local translations of space-time of the Poincare group *P*.

**Definition**. Let $\alpha : g \to \alpha_g$ be a group representation of the translation subgroup *T* of the Poincare group as a discrete group acting on the von Neumann algebra *O(D)*. A representation $\pi$ of *O(D)* acting on a Hilbert space *H* is covariant if there is a homomorphism $g \to U_g$ from *G* to the group of unitary operators on *H* with $\pi(\alpha_g(A)) = U_g \pi(A) U_g^* \quad \forall A \in O(D)$.

If $\varphi$ is a normal state of *O(D)* then $\varphi \circ \alpha_g$ is also a normal state since each automorphism preserves the algebraic structure and hence preserves complete additivity. If *S* denotes the set of all normal states of *O(D)* then the direct sum $\pi = \oplus \{\pi_\varphi ; \varphi \in S\}$ of their Gelfand-Naimark-Segal (GNS) representations is a faithful representation of *O(D)* as a von Neumann algebra acting on a Hilbert space *H* which is the direct sum of the GNS Hilbert spaces. If we define

$U_g \left( \oplus_{\varphi \in S} \pi_\varphi(A_\varphi) x_\varphi \right) = \oplus_{\varphi \in S} \pi_\varphi(\alpha_g(A_{\varphi \circ \alpha_g})) x_\varphi$ as a mapping on each of the pre-Hilbert spaces for the GNS constructions, then $U_g$ extends to a unitary operator on *H* and the representation $\pi = \oplus \{\pi_\varphi ; \varphi \in S\}$ is a faithful normal representation of *O(D)*. We can therefore assume that *T* acting on *O(D)* as a discrete group of automorphisms is unitarily implemented, if necessary.

**Definition.** (Moffat, Stormer). Let $\alpha : g \to \alpha_g$ be a group representation of the translation subgroup *T* of the Poincare group as a discrete group acting on the von Neumann algebra *O(D)*. If *E* and *F* are projections in *O(D)* we say that *E* and *F* are *T*-equivalent if there is a set of operators $\{A_g ; g \in T, A_g \in O(D)\}$ with $E = \sum_g A_g^* A_g$ and $F = \sum_g \alpha_g(A_g A_g^*)$.

**Definition.** We write this *T*-equivalence as $E \approx F$ and call it a *T-twisted equivalence.* In the special case that each $A_g$ is a projection, this definition is a direct non-commutative generalisation of Hopf equivalence.

**Definition** (Moffat, Stormer). A projection *F* is defined to be *T-finite* if *F* contains no proper sub-projections which are *T*-equivalent to *F*. The algebra *O(D)* is defined to be *T-finite, or T-Type II(1),* if the identity of *O(D)* is a *T*-finite projection. *O(D)* is *T-semifinite, or T-Type II(∞)* if every projection in O(D) dominates a *T*-finite projection. *O(D)* is *T-purely infinite, or T-Type III,* if *O(D)* does not contain any *T*-finite projections.

The *T*-type III case is the most difficult to analyse. In the *T*-type III case there is not even the 'shadow' of a trace. A *T*-invariant trace is a bounded faithful normal linear mapping *τ: O(D)*→$\mathbb{C}$ with;

$$\tau(AB) = \tau(\alpha_g(AB)) = \tau(BA) \quad \forall g \in T; A, B \in O(D).$$

If $\tau$ is a trace, then by the earlier remarks we can assume that the group representation of $T$, as a discrete group, is unitarily implemented by the unitary representation $U: g \to U_g$ so that $\tau$ is automatically $T$-invariant. Further, if $F$ is a $T$-finite projection and $E \sim F$ in the sense of Murray and von Neumann then $E \leq F$ and $E \sim F$ imply that $E \leq F$ and $E \approx F$ (using only the identity of the group). Thus $E = F$ and $F$ being $T$-finite implies $F$ is finite.

(Stormer, 1973) established that $O(D)$ is $T$-semifinite if and only if there is a faithful normal semifinite $T$-invariant trace on $O(D)$.

*The Crossed Product Algebra of O(D)*

Assume (by taking a faithful representation if necessary) that $O(D)$ acts on a Hilbert space $H$. Define the Dirac function $\varepsilon_g$ to take the value 1 at g and zero elsewhere on $T$. Then $\{\varepsilon_g; g \in T\}$ is an orthonormal basis for the Hilbert space $l^2(T)$. Given $l^2(T)$ and H we can form the tensor product Hilbert space $H \otimes l^2(T)$. Define;

$$U_h(x \otimes \varepsilon_g) = x \otimes \varepsilon_{gh^{-1}} \text{ for } x \in H,\ g, h \in T$$
$$\Phi(A)(x \otimes \varepsilon_g) = \alpha_g(A) x \otimes \varepsilon_g \text{ for } A \in O(D), g \in T$$

Then $U_h$ extends to a unitary operator on $H \otimes l^2(T)$ and the mapping $h \to U_h$ is a group homomorphism from the translation group T into the group of unitaries acting on $H \otimes l^2(T)$.

Similarly $\Phi(A)$ extends to a bounded linear operator on $H \otimes l^2(T)$ for all $A$ in $O(D)$ and the mapping $h \to U_h$ implements the automorphic representation $h \to \alpha_h$.

The transformation $\Phi$ is an ultraweakly continuous *isomorphism of $O(D)$ and it follows that $\Phi(O(D))$ is a von Neumann algebra. Finite sums $\sum_j U_{g_j} \Phi(A_j)$ form a *algebra denoted $(O(D) \times T)_0$ which contains $\Phi(O(D))$. The cross product algebra $O(D) \times T$ is defined as the closure of the *algebra $(O(D) \times T)_0$ for the ultraweak operator topology. The crossed product algebra $O(D) \times T$ can be used to prove the following structural result.

Assume *O(D1)* and *O(D2)* are local von Neumann algebras in space-time regions *D1* and *D2* which are not space-like separated. Let *G* and *H* be discrete representations of the translation subgroup of the Poincare group as automorphisms of *O(D1)* and *O(D2)* respectively. Then if either *O(D1)* or *O(D2)* is *G/H*-purely infinite (*G/H*-Type III), the joint algebra $O(D1) \otimes O(D2)$ is $G \times H$- purely infinite (equivalently $G \times H$-type III) under the action of the joint representation G×H of the translation group. If both O(D1) and O(D2) are G/H finite or G/H semifinite, then the same applies to the joint algebra $O(D1) \otimes O(D2)$. These results follow from the fact that (Moffat, 1974); $(O(D1) \times G) \otimes O(D2) \times H)$ is spatially *isomorphic to (OD1⊗OD2)×(G×H).

**A Symmetry of Types**

In this part of our analysis of the structure of *O(D)* we find a pleasing symmetry for purely infinite type III algebras between the *T*-type of *O(D)* and the corresponding Murray-von Neumann type of its cross product algebra *O(D)×T*. First we have to prove the following key result. Recall that, by construction, the crossed product von Neumann algebra O(D) ×T contains the embedded closed sub-algebra *Φ(O(D))*, isomorphic to *O(D)*.

**Theorem 4.** There is an ultraweakly continuous mapping, denoted Γ, from O(D) ×T to O(D) such that the restriction $\Gamma|_{\Phi(O(D))} = \Phi^{-1}$, the inverse of the embedding of the algebra O(D); and the composite map $\Gamma \circ \Phi : O(D) \times T \to \Phi(O(D))$ is a continuous projection of norm one.

**Proof.** Continuing with the notation introduced earlier; the map $x \to x \otimes \varepsilon_g : H \to H_g$, where $H_g$ is a closed linear subspace of $H \otimes l^2(T)$, is both isometric and linear. Since the set $\{\varepsilon_g ; g \in T\}$ is orthonormal, the Hilbert space $K = H \otimes l^2(T)$ is the direct sum of the $H_g$'s and every element x of K can be represented as $x = \sum_{g \in T} x_g \otimes \varepsilon_g$ with $\|x\|^2 = \sum_{g \in T} \|x_g\|^2 < \infty$.

If $E_g$ is the projection from K onto $H_g$, and $B = \sum_g U_g \Phi(A_g)$ is an element of $(O(D) \times T)_0$ then straightforward arguments show that $E_s B E_t = E_s U_{s^{-1}t} \Phi(A_{s^{-1}t}) E_t$. Taking the weak closure, we have $B \in O(D) \times T$ with $B = \lim_\alpha \{B^\alpha ; E_s B^\alpha E_t = E_s U_{s^{-1}t} \Phi(D^\alpha_{s^{-1}t}) E_t\}$. From the Kaplansky density theorem (Kadison and Ringrose, 1983) we can choose $B^\alpha$ with $\|B^\alpha\| \leq \|B\|$ and the net $D^{\alpha}_{s^{-1}t}$ is then a bounded net in the ultraweakly compact ball of radius $\|B\|$. It thus has a subnet converging to an element $D_{s^{-1}t}$ of $O(D)$. From this we have the following expression;

$$E_s B E_t = E_s U_{s^{-1}t} \Phi(D_{s^{-1}t}) E_t \quad\quad\quad\quad\quad\quad\quad (1)$$

In particular we have $E_e B E_e = E_e \Phi(D_e) E_e$. If we define $\Gamma(B) = D_e$ then clearly the mapping $\Gamma$ is linear, and $\Gamma|_{\Phi(O(D))} = \Phi^{-1}$. Finally if $B^{\alpha} \to B$ ultraweakly then $E_e B^{\alpha} E_e \to E_e B E_e$ and the mapping $\Gamma$ is ultraweakly continuous.

If $B$ is in the kernel of $\Gamma$ then $D_e = 0$. From equation (1) above this implies that $E_s B E_s = 0 \;\; \forall s$ and thus $B = 0$; the kernel of $\Gamma$ is $\{0\}$ and $\Gamma$ is a faithful mapping. This shows that $\Gamma$ has the required properties, and completes the proof.

This allows us to now prove the following key structural result.

**Theorem 5**. *O(D)* is *T*-type III if and only if the crossed product algebra *O(D)* ×*T* is type III in the sense of Murray-von Neumann.

**Proof.** If *O(D)* is not *T*-type III then it contains a non-trivial *T*-finite projection E. Then it follows that if Φ is the identification of *O(D)* within the crossed product algebra *O(D)*×*T* then Φ(E) is finite in the sense of Murray-von Neumann. Thus *O(D)*×*T* is not type III.

Conversely assume the crossed product algebra *O(D)*×*T* is not type III. From Theorem 4 we know that there is a faithful normal projection Γ of norm one from *O(D)*×*T* onto *O(D)*. From (Sakai, 2012) it follows that *O(D)* cannot be type III. Thus *O(D)Z* is semifinite for some projection Z in the centre of *O(D)*. From (Stormer, 1970), *O(D)Z* is *T*-semifinite thus *O(D)* cannot be *T*-type III. This completes the proof.

Takesaki, Masamichi. "On the singularity of a positive linear functional on operator algebra." *Proceedings of the Japan Academy* 35.7 1959: 365-366.